\documentclass[10pt,twocolumn,twoside]{IEEEtran}

\usepackage{amsmath}
\usepackage{amssymb}
\usepackage{latexsym}
\usepackage{verbatim}
\usepackage{subfigure}
\usepackage[final]{graphicx}
\usepackage{psfrag}

\newtheorem{theorem}{Theorem}
\newtheorem{proposition}{Proposition}
\newtheorem{lemma}{Lemma}

\newtheorem{remark}{Remark}

\newtheorem{example}{Example}

{\begin{list}               %    with flush left bullets.
    {$\bullet$ \hfill}{
        \setlength{\leftmargin}{\parindent}
        \setlength{\parsep}{0.04\baselineskip}
        \setlength{\itemsep}{0.5\parsep}
        \setlength{\labelwidth}{\leftmargin}
        \setlength{\labelsep}{0em}}
    }
{\end{list}}

\providecommand{\eref}[1]{\eqref{eq:#1}}  % call \eqref from amstex
\providecommand{\cref}[1]{Chapter~\ref{chap:#1}}
\providecommand{\sref}[1]{Section~\ref{sec:#1}}
\providecommand{\fref}[1]{Figure~\ref{fig:#1}}

\providecommand{\R}{\ensuremath{\mathbb{R}}}
\providecommand{\C}{\ensuremath{\mathbb{C}}}

\providecommand{\abs}[1]{\lvert#1\rvert}
\providecommand{\norm}[1]{\lVert#1\rVert}
\providecommand{\inprod}[1]{\langle#1\rangle}
\providecommand{\set}[1]{\left\{#1\right\}}

\providecommand{\bydef}{\overset{\text{def}}{=}}

\renewcommand{\vec}[1]{\ensuremath{\boldsymbol{#1}}}
\providecommand{\mat}[1]{\ensuremath{\boldsymbol{#1}}}

\providecommand{\mD}{\mat{D}}

\providecommand{\mI}{\mat{I}}

\providecommand{\va}{\vec{a}}

\providecommand{\vx}{\vec{x}}

\usepackage{cite}

\usepackage{color}

\usepackage{booktabs,multirow}

\usepackage{enumitem}
\usepackage{commath}

\usepackage{algorithm, algorithmic}

\usepackage{hyperref}

\usepackage{esdiff}

\usepackage{pgf,tikz,psfrag}

\usepackage{dsfont}
\newcommand{\charfn}{\mathds{1}}

\providecommand{\lm}{\lambda}

\providecommand{\muf}{\phi}
\providecommand{\uf}{\psi}

\providecommand{\ub}{\tau}

\providecommand{\alpw}{\alpha_\mathrm{weak}}
\providecommand{\ropt}{\rho_\mathrm{optimal}}
\providecommand{\rhoaT}{\rho(\alpha; \mathcal{T}(\cdot))}

\providecommand{\T}{\mathcal{T}(\cdot)}
\providecommand{\Ty}{\mathcal{T}(y)}
\providecommand{\TY}{\mathcal{T}(Y)}
\providecommand{\Tmm}{\mathcal{T}_\mathrm{MM}(\cdot)}
\providecommand{\Tmmy}{\mathcal{T}_\mathrm{MM}(y)}
\providecommand{\Topt}{\mathcal{T}_\mathrm{optimal}(\cdot)}
\providecommand{\Topty}{\mathcal{T}_\mathrm{optimal}(y)}

\providecommand{\Ttrimy}{\mathcal{T}_\mathrm{trim}(y)}

\providecommand{\Tsuby}{\mathcal{T}_\mathrm{subset}(y)}
\providecommand{\Tast}{\mathcal{T}^\ast(\cdot)}
\providecommand{\Tasty}{\mathcal{T}^\ast(y)}
\providecommand{\Tae}{\mathcal{T}^{\varepsilon}_\alpha(\cdot)}
\providecommand{\Taey}{\mathcal{T}^{\varepsilon}_\alpha(y)}

\providecommand{\caey}{c^{\varepsilon}_\alpha(y)}
\providecommand{\vae}{v^{\varepsilon}_\alpha}

\providecommand{\Sy}{\Gamma_Y}
\providecommand{\Qc}{\mathcal{Q}(c(\cdot),\beta)}
\providecommand{\Lc}{\mathcal{L}(c(\cdot))}

\providecommand{\Lcv}{\mathcal{L}(c_v(\cdot))}

\providecommand{\vxi}{\vec{\xi}}

\begin{document}
%.
% Paper Title
\title{Optimal Spectral Initialization for Signal \\Recovery with Applications to Phase Retrieval}

\author{%
Wangyu Luo, Wael Alghamdi and Yue M. Lu
\thanks{The authors are with the John A. Paulson School of Engineering and Applied Sciences, Harvard University, Cambridge, MA 02138, USA. This work was supported by the US National Science Foundation under grant CCF-1718698.}
}

\markboth{} {Luo \MakeLowercase{\textit{et al.}}: Optimal Spectral Initialization for Signal Recovery}

\maketitle

\begin{abstract}
We present the optimal design of a spectral method widely used to initialize nonconvex optimization algorithms for solving phase retrieval and other signal recovery problems. Our work leverages recent results that provide an exact characterization of the performance of the spectral method in the high-dimensional limit. This characterization allows us to map the task of optimal design to a constrained optimization problem in a weighted $L^2$ function space. The latter has a closed-form solution. Interestingly, under a mild technical condition, our results show that there exists a fixed design that is uniformly optimal over all sampling ratios. Numerical simulations demonstrate the performance improvement brought by the proposed optimal design over existing constructions in the literature. In a recent work, Mondelli and Montanari have shown the existence of a weak reconstruction threshold below which the spectral method cannot provide useful estimates. Our results serve to complement that work by deriving the fundamental limit of the spectral method beyond the aforementioned threshold.
\end{abstract}


%!TEX root = optimal_spectral.tex

\section{Introduction}

An active line of recent work studies nonconvex optimization algorithms for solving the classical phase retrieval problem (see, \emph{e.g.}, \cite{Netrapalli:2013, Candes:2015, Chen:2015, LiGL:15, zhang2016provable, WangGY:2016, ChiL:16, ma2018optimization}). Compared to methods using convex relaxation \cite{Candes:2013xy, Candes:2014ty, Jaganathan:2013zl, Waldspurger:2015rz}, the nonconvex approaches tend to require much lower computational complexity and memory footprints. A key ingredient in many such algorithms is a simple yet highly effective spectral method \cite{Li:92, Netrapalli:2013, Chen:2015}. It provides an initial estimate that is sufficiently close to the target signal. Starting from this ``warm start'', local search schemes such as gradient descent can then carry out further refinement to reach globally optimal solutions.

This paper studies the optimal design of the aforementioned spectral method. Throughout the paper, we consider the following sensing model. Let $\vxi \in \C^n$ denote the target signal we seek to estimate, and $\set{\va_i \in \C^n}_{1 \le i \le m}$ a collection of sensing vectors. Given $s_i = \inprod{\va_i, \vxi}$, the $i$th measurement $y_i$ is drawn independently from
\begin{equation}\label{eq:model}
y_i \sim p\big(y \, \big\vert \, \abs{s_i}\big),\qquad 1 \le i \le m,
\end{equation}
where $p(\cdot \, \vert \, \cdot)$ is a conditional density function modeling the (potentially noisy) sensing process. Clearly, the phase information of $s_i$ is missing, as $y_i$ only depends on the magnitude of $s_i$. The spectral method we study consists of two simple steps \cite{Li:92, Netrapalli:2013, Chen:2015}. First, construct a data matrix as
\begin{equation}\label{eq:D_mtx}
\mD \bydef \frac{1}{m} \sum_{i=1}^m \mathcal{T}(y_i) \va_i \va_i^\ast,
\end{equation}
where $\mathcal{T}: \R \mapsto \R$ is a user-specified \emph{preprocessing function}. Second, we compute $\vx_1$, an eigenvector associated with the largest eigenvalue of $\mD$. The vector $\vx_1$ is then our initial estimate of $\vxi$ (up to an unknown scalar). 

The idea of this spectral method first appeared in the statistics literature under the name of principal Hessian directions \cite{Li:92}. In the context of phase retrieval, it was introduced by Netrapalli, Jain, and Sanghavi as an initialization step for their alternating minimization algorithm \cite{Netrapalli:2013}. Finite sample performance analysis of the spectral method can be found in \cite{Netrapalli:2013, Candes:2015, Chen:2015}. Under Gaussian design, the normalized correlation between the eigenvector $\vx_1$ and the target vector $\vxi$ is shown to approach $1$ with high probability, provided that the number of samples $m$ is sufficiently large with respect to the signal dimension $n$. In particular, by introducing a trimming step on the measurements (see \eref{trimming} below), Chen and Candes \cite{Chen:2015} show that it suffices to have $m \ge c \, n$, where $c$ is some sufficiently large constant.

In \cite{LuL:18}, Lu and Li presented an asymptotically exact characterization of the performance of the spectral methods. Specifically, under Gaussian design and when $m, n \to \infty$ at a fixed ratio $\alpha = m/n$, they show that the normalized correlation between $\vxi$ and $\vx_1$ converges in probability to a deterministic value, \emph{i.e.},
\begin{equation}\label{eq:limit}
\frac{\abs{\inprod{\vxi, \vx_1}}^2}{\norm{\vxi}^2\norm{\vx_1}^2} \xrightarrow[n \to \infty]{\mathcal{P}} \rhoaT.
\end{equation}
Moreover, explicit formulas are available to compute the limit value $\rhoaT$. [See \sref{asymptotic} for details.] The above asymptotic characterization was first derived for the real-valued case and under the assumption that $\Ty \ge 0$ \cite{LuL:18}. Then Mondelli and Montanari generalize the characterization to the complex-valued case in \cite{MondelliM:17}, where the assumption that $\Ty$ be nonnegative is also shown to be unnecessary.

%In addition to providing an exact characterization of the performance of the spectral method, the above asymptotic prediction also reveals a phase transition phenomenon that is associated with the sampling ratio $\alpha$. In particular, there exists a threshold value, denoted by $\alpha_{c}(\mathcal{T})$, that marks the transitions between two very different phases. (a) An \emph{uncorrelated phase} takes place when the sampling ratio $\alpha_{c}(\mathcal{T})$. Within this phase, the limiting value $\rho(\alpha) = 0$, meaning that the estimate from the spectral method is asymptotically uncorrelated with the target vector $\vxi$. (b) A \emph{correlated phase} takes place when $\alpha > \alpha_{c}(\mathcal{T})$. Within this phase, the limiting value $\rho(\alpha) > 0$. 

The performance of the spectral method depends heavily on the form of the preprocessing function $\T$ used in \eref{D_mtx}. (Accordingly, on the right-hand side of \eref{limit}, our notation for the limiting squared correlation $\rhoaT$ makes its dependence on $\T$ explicit.) Several designs have been proposed in the literature, including the trimming scheme introduced in \cite{Chen:2015}:
\begin{equation}\label{eq:trimming}
\Ttrimy = y \, \charfn_{\set{\abs{y} \le a}},
\end{equation}
and the subset scheme proposed in \cite{WangGY:2016}:
\begin{equation}\label{eq:subset}
\Tsuby = \charfn_{\set{\abs{y} \ge b}}.
\end{equation}
In \eref{trimming} and \eref{subset}, $\charfn_{\set{\cdot}}$ denotes the indicator function on a set, and $a, b$ are some tuning parameters. See also \cite{zhang2016provable} for yet another design that improves the robustness of the method.

%where $a > 0$ is some fixed parameter and $\charfn_{\set{\cdot}}$ denotes the indicator function. A different design can be found in \cite{}, where the authors choose
%for some fixed $b > 0$. We note that the preprocessing done in \eref{trimming} and \eref{subset} confines the values of $\mathcal{T}(y)$ to a bounded interval. As shown in \cite{}, this boundedness property is crucial in allowing one to achieve order-wise optima sample complexity.

While the existing designs in the literature are all based on sound intuitions (and ingenuity), they are not expected to be optimal. Equipped with the exact asymptotic characterizations obtained in \cite{LuL:18, MondelliM:17}, we now have the luxury to ask the following question: given any specific sensing model in \eref{model},  what is the corresponding \emph{optimal form} of the preprocessing function? Specifically, we consider the following optimal design problem:
\begin{equation}\label{eq:opt_design_prob}
\ropt(\alpha) \bydef \sup_{\T \in \mathcal{F}}\, {\rhoaT},
\end{equation}
where $\rhoaT$ is the limiting squared correlation in \eref{limit}, and $\mathcal{F}$ denotes a set of feasible functions from which we search for the optimal one. The exact definition of $\mathcal{F}$ will be given in \eref{feasible} in \sref{results}. It serves to restrict the search space to make sure that the asymptotic predictions obtained in \cite{LuL:18, MondelliM:17} are applicable. In what follows, we refer to $\ropt(\alpha)$ as the \emph{optimal performance curve}.

The first result addressing the optimal design problem was obtained by Mondelli and Montanari \cite{MondelliM:17}, who show that
\begin{equation}\label{eq:ropt_0}
\ropt(\alpha)= 0, \qquad \text{for } \alpha \le \alpw,
\end{equation}
where $\alpw$ is called the \emph{weak reconstruction threshold} in their paper. Given the definition of $\ropt(\alpha)$, the result in \eref{ropt_0} implies that, when the sampling ratio $\alpha \le \alpw$, the spectral method cannot provide an estimate that has nontrivial correlation with the target vector $\vxi$, no matter how one chooses the preprocessing function. Moreover, Mondelli and Montanari show that
\begin{equation}\label{eq:ropt_1}
\ropt(\alpha)> 0, \qquad \text{for } \alpha > \alpw.
\end{equation}
They establish this by constructing a specific preprocessing function, denoted by $\Tmm$, such that
\[
\rho(\alpha; \Tmm) > 0 \qquad \text{for } \alpha > \alpw.
\]
For any sensing model \eref{model}, explicit formulas are provided in \cite{MondelliM:17} to compute $\alpw$ and $\Tmm$. We defer such technical details to \sref{results} [see \eref{alpw} and \eref{Tmm}]. 

We note that, while the preprocessing functions $\Tmm$ serve to show that the optimal performance curve $\ropt(\alpha)$ is strictly positive when $\alpha > \alpw$, these functions do not solve the optimization problem in \eref{opt_design_prob}. Thus, important questions remain as to what $\ropt(\alpha)$ should be beyond $\alpw$ and whether there are optimal functions that can potentially achieve this bound.

In this paper, we present a complete solution of the optimal design problem formulated in \eref{opt_design_prob}. Specifically, we provide an exact analytical expression for $\ropt(\alpha)$ for all $\alpha > 0$ and for any sensing model. Moreover, under a mild technical condition (which is satisfied by many sensing models), we construct an optimal preprocessing function $\Topt$ that solves the design problem. Somewhat surprising about the optimal solution is the fact that $\Topt$ does not depend on the sampling ratio $\alpha$. In other words, the proposed $\Topt$ is \emph{uniformly optimal} for all $\alpha$. Finally, when the aforementioned technical condition does not hold, we show that the supremum in \eref{opt_design_prob} cannot be achieved by any function in the feasible set $\mathcal{F}$. In this case, we construct a family of preprocessing functions $\Tae$ whose performance will approach $\ropt(\alpha)$ as $\varepsilon \to 0$. 

The rest of the paper is organized as follows. Our main results are stated as Theorem~\ref{thm:optimal} in \sref{results}. To illustrate these results, we present worked examples corresponding to two different sensing models. Numerical simulations demonstrate the performance improvements brought by the proposed optimal design over heuristic choices given in \eref{trimming} and \eref{subset} as well as the functions $\Tmm$ constructed in \cite{MondelliM:17}. To set the stage for proving our results, \sref{technical} recalls the asymptotic characterization of the spectral method obtained in previous work \cite{LuL:18, MondelliM:17}. This characterization allows us to map the optimal design problem in \eref{opt_design_prob} to a (constrained) optimization problem in a weighted $L^2$ function space. The proof of Theorem~\ref{thm:optimal} is given in \sref{proof}. Although we state and prove our results for the more general complex-valued case in this paper, the treatment of the real-valued case is the same, \emph{mutatis mutandis}. See Remark~\ref{rem:real} in \sref{results} for an explanation of these changes.

%!TEX root = optimal_spectral.tex

\section{Main Results}
\label{sec:results}

We start by introducing two functions that will play central roles in our later technical discussions. Let 
\begin{equation}\label{eq:S_rv}
S \sim \mathcal{CN}(0, 1) 
\end{equation}
be a complex-valued standard normal random variable. Define
\begin{equation}\label{eq:eta}
\eta(y) \bydef \mathbb{E}_S \big[p\big(y \; \big\vert \; \abs{S}\big)\big]
\end{equation}
and
\begin{equation}\label{eq:mu}
\mu(y) \bydef \mathbb{E}_S \big[\abs{S}^2 p\big(y \, \big\vert \, \abs{S}\big)\big],
\end{equation}
where $p(\cdot \vert \cdot)$ is the conditional density function associated with the sensing model in \eref{model}. It is easy to verify that
\begin{equation}\label{eq:unit_int}
\int \eta(y) \dif y = \int \mu(y) \dif y = 1.
\end{equation}

The two functions $\eta(y)$ and $\mu(y)$ allow us to conveniently state the results of \cite{MondelliM:17} as well as our new results. For example, the weak reconstruction threshold $\alpw$ introduced in \cite{MondelliM:17} can be written as
\begin{equation}\label{eq:alpw}
\alpw \bydef \Big[\int \frac{[\mu(y) - \eta(y)]^2}{\eta(y)} \dif y\Big]^{-1}.
\end{equation}
Moreover, the preprocessing function constructed in \cite{MondelliM:17} is
\begin{equation} \label{eq:Tmm}
\Tmmy \bydef \frac{\sqrt{\alpw}\, \Tast}{\sqrt{\alpha} - (\sqrt{\alpha} - \sqrt{\alpw}) \Tast},
\end{equation}
for $\alpha > \alpw$, where
\begin{equation}\label{eq:T_ast}
\Tasty \bydef 1 - \eta(y) / \mu(y).
\end{equation}

In Appendix~\ref{appendix:alpw}, we show that the integral on the right-hand side of \eref{alpw} is always well-defined. Moreover, $\alpw \ge 1$ under any sensing model, with the lower bound achieved by the case of noiseless phase retrieval, \emph{i.e.}, $y_i = \abs{\inprod{\va_i, \vxi}}^2$. 

\subsection{Optimal Design}

Our optimal design of the preprocessing function leverages upon the asymptotic characterizations given in \cite{LuL:18, MondelliM:17}, which are derived under some technical assumptions on $\T$. Specifically, let $Y$ be a random variable whose conditional distribution given $S$ is 
\begin{equation}\label{eq:Y_rv}
Y \vert S \sim p\big(y \,\big\vert\, \abs{S}\big).
\end{equation}
Let $\Sy$ denote the support of the probability measure of $Y$. We shall assume that the preprocessing function $\mathcal{T}(y)$ belongs to the following \emph{feasible set}:
\begin{equation}\label{eq:feasible}
\mathcal{F} \bydef \Big\{\mathcal{T}(y): 0 < \sup_{y \in \Sy} \mathcal{T}(y) < \infty \text{ and } \inf_{y \in \Sy} \mathcal{T}(y) > -\infty\Big\}.
\end{equation}
In words, we require that $\mathcal{T}(y)$ should have a bounded range and that the upper boundary of that range should be positive.

\begin{theorem}\label{thm:optimal}
Suppose that the target signal $\vxi$ is an arbitrary vector in $\C^n$ with $\norm{\vxi} = 1$, and that the sensing vectors $\set{\va_i}_{1 \le i \le m}$ are drawn independently from the rotationally symmetric complex Gaussian distribution, \emph{i.e.}, $\va_i \sim_\text{i.i.d.} \mathcal{CN}(\vec{0}, \mI_n)$. As $m, n \to \infty$ with $m/n \to \alpha \in (0, \infty)$, the following hold with respect to the optimal design problem in \eref{opt_design_prob}:
\begin{enumerate}
\item For each $\alpha > \alpw$, let $\beta_\alpha$ denote the unique positive root of the equation $f(\beta) = 1/\alpha$, where
\begin{equation}\label{eq:fbeta}
f(\beta) \bydef \int_{\Sy} \frac{[\mu(y) - \eta(y)]^2}{\eta(y) + \mu(y)/\beta} \dif y.
\end{equation}
Then
\begin{equation}\label{eq:rho_opt_beta}
\ropt(\alpha) = \begin{cases}
(1 + \beta_\alpha)^{-1}, &\text{for } \alpha > \alpw;\\
0, &\text{otherwise}.
\end{cases},
\end{equation}
\item If $\inf_{y \in \Sy} \frac{\mu(y)}{\eta(y)} > 0$, the optimal performance curve $\ropt(\alpha)$ can be achieved by 
\begin{equation}\label{eq:T_uniform}
\Topty = 1 - {\eta(y)}/{\mu(y)} \in \mathcal{F}.
\end{equation}

\item If $\inf_{y \in \Sy} \frac{\mu(y)}{\eta(y)} = 0$, then $\ropt(\alpha)$ cannot be achieved by any function $\T \in \mathcal{F}$. However, there exists a family of functions $\set{\Tae \in \mathcal{F}}_{0 < \varepsilon < 1} $ such that
\[
\lim_{\varepsilon \to 0} \,\rho(\alpha; \Tae) = \rho_\text{optimal}(\alpha).
\]
As an explicit construction of such a family, we can set $\Taey=\frac{c^\varepsilon_\alpha(y)}{1+c^\varepsilon_\alpha(y)}$, where
\begin{equation}\label{eq:c_ast_e}
c^\varepsilon_\alpha(y) = \max\left\{v_\alpha^\varepsilon \frac{\mu(y) - \eta(y)}{\eta(y) + \mu(y)/\beta_\alpha}, -1+\varepsilon \right\}.
\end{equation}
Here, $\beta_\alpha$ is the same constant as in \eref{rho_opt_beta}, and  $v_\alpha^\varepsilon \ge 1$ is a scalar that can be uniquely determined by the linear constraint $\int_{\Sy} c^\varepsilon_\alpha(y)[\mu(y)-\eta(y)]\dif y = 1/\alpha$.
\end{enumerate}
\end{theorem}

\begin{remark}
Theorem~\ref{thm:optimal}, whose proof is given in \sref{proof}, provides a complete solution to the optimal design problem formulated in \eref{opt_design_prob}. As mentioned earlier, when $\inf_{y \in \Sy} \frac{\mu(y)}{\eta(y)} > 0$, the preprocessing function $\Topt$ given in \eref{T_uniform} is uniformly optimal, as it does not depend on the sampling ratio $\alpha$. We also note that there is a strong connection between $\Topt$ and the function $\Tmm$ designed in \cite{MondelliM:17}. In fact, $\Topt$ is exactly equal to $\Tast$ in \eref{T_ast}.
\end{remark}

\begin{remark}[The real-valued case]\label{rem:real}
The results of Theorem~\ref{thm:optimal} can be directly applied to the real-valued case, after we make the following changes: (1) In the definitions of $\eta(y)$ and $\mu(y)$ in \eref{eta} and \eref{mu}, the random variable $S$ is now drawn from $\mathcal{N}(0, 1)$ instead of $\mathcal{CN}(0, 1)$; (2) In the statement of Theorem~\ref{thm:optimal}, we shall assume $\vxi \in \R^n$ and that the sensing vectors $\va_i \sim_\text{i.i.d.} \mathcal{N}(\vec{0}, \mI_n)$.
\end{remark}

% For the special case of noiseless phase retrieval, \emph{i.e.}, $y_i = \abs{\inprod{\va_i, \vxi}}^2$, the weak reconstruction threshold is equal to $1$ (resp. $1/2$) for the complex-valued (resp. real-valued) case. The corresponding optimized preprocessing function is given as
%\[
%\mathcal{T}_\mathrm{MM}(y) = \frac{\sqrt{\alpw} (1-1/y)}{\sqrt{\alpha} - (\sqrt{\alpha} - \sqrt{\alpw}) (1-1/y)}, \ \text{for } \alpha > \alpha_\mathrm{weak}.
%\]

\subsection{Worked Examples}
\label{sec:examples}

To show how the results stated in Theorem~\ref{thm:optimal} can be applied in practice, we present two worked examples corresponding to two different sensing models.

\begin{example}[Poisson measurements]
Here we consider the Poisson model, where
\[
y_i \sim \text{Poisson}\big( \kappa \cdot \abs{\inprod{\va_i, \vxi}}^2\big),
\]
and  $\kappa > 0$ is an additional parameter indicating the signal-to-noise ratio in the sensing process. Note that the measurements $\set{y_i}$ here are nonnegative integers instead of continuous variables. Theorem~\ref{thm:optimal} still applies. We just need to treat the integration in \eref{fbeta} as summations.

Let $Z = \abs{S}^2$, with $S$ defined as in \eref{S_rv}. It is well-known that $Z$ follows the exponential distribution with parameter $1$. Using this property, we can compute the function in \eref{eta} as
\begin{align}
\eta(y)&=  \mathbb{E}_Z \bigg[\frac{e^{-\kappa Z} (\kappa Z)^y}{y!}\bigg] \nonumber \\
&=\frac{\kappa^y}{y!}\int_0^\infty e^{-\kappa z} z^y e^{-z} \dif z =\frac{\kappa^y}{(\kappa+1)^{y+1}}.\label{eq:eta_Poisson}
\end{align}
Similarly, the function in \eref{mu} becomes
\begin{equation}\label{eq:mu_Poisson}
\mu(y)=\frac{\kappa^y(y+1)}{(\kappa+1)^{y+2}}.
\end{equation}
Since $y$ only takes nonnegative integer values,
\[
\inf_{y} \, \frac{\mu(y)}{\eta(y)}=\inf_y\, \frac{y+1}{\kappa+1}=\frac{1}{\kappa+1}>0.
\]
It then follows from Theorem~\ref{thm:optimal} that there exists a uniformly optimal preprocessing function, which in our case is
\begin{equation}\label{eq:T_Poiss}
\Topty =1-\frac{\kappa+1}{y+1}=\frac{y-\kappa}{y+1}.
\end{equation}

Substituting \eref{eta_Poisson} and \eref{mu_Poisson} into \eref{alpw}, we get
\[
\begin{aligned}
\alpw &= \bigg[\sum_{y=0}^\infty \frac{[\mu(y) - \eta(y)]^2}{\eta(y)}\bigg]^{-1}\\
&=\bigg[\sum_{y=0}^\infty \frac{\kappa^y (y - \kappa)^2}{(\kappa+1)^{y+3}} \bigg]^{-1}=1+{1}/{\kappa}.
\end{aligned}
\]
Finally, the function $f(\beta)$ in \eref{fbeta} can be calculated as
\begin{align}
f(\beta)=&  \frac{\beta}{(\kappa+1)^2}\sum_{y=0}^\infty\left( \frac{\kappa}{\kappa+1}\right)^{y}\frac{(y-\kappa)^2}{y+\beta(\kappa+1)+1} \nonumber\\
=&C_{\beta, \kappa}\int_{0}^{\frac{\kappa}{\kappa+1}}\frac{x^{\beta(\kappa+1)}}{1-x}\dif x-\beta(\beta+1),\label{eq:fbeta_Poisson}
\end{align}
where $C_{\beta, \kappa} = (\beta+1)^2\beta (1 + \tfrac{1}{\kappa})^{\beta(\kappa+1)+1}$. In reaching \eref{fbeta_Poisson} we have used the identity that
\[
\int_{0}^{v}\frac{x^u}{1-x}\dif x = \sum_{y=0}^\infty \frac{v^{y+u+1}}{y + u + 1},
\]
for any $u > 0$ and $0 < v < 1$. In our case, we choose $u = \beta(\kappa+1)$ and $v = {\kappa}/({\kappa+1})$.

 \fref{fbeta} shows the function $f(\beta)$ for $\kappa = 5$. We can further show that this function is strictly increasing, with $\lim_{\beta \to 0^+} f(\beta) = 0$ and $\lim_{\beta \to \infty} f(\beta) = 1/\alpw$. It follows that, for each $\alpha > \alpw$, there is a unique $\beta_\alpha > 0$ satisfying the equation $f(\beta_\alpha) = 1/\alpha$. Applying Theorem~\ref{thm:optimal}, the optimal performance curve is simply $\ropt(\alpha) = (1 + \beta_\alpha)^{-1}$.

In \fref{Poisson}, we compare the proposed optimal preprocessing function in \eref{T_Poiss} against the the trimming scheme in \eref{trimming}, the subset scheme in \eref{subset}, as well as $\Tmm$ in \eref{Tmm}. In our experiments, the signal dimension is set to $n = 4096$ and $\kappa=5$. For each given $\alpha$, we set the parameter $a$ in \eref{trimming} to be the optimal integer choice within $\set{1, 2, \ldots, 50}$. For \eref{subset}, its parameter $b$ is tuned in the same way. For our optimal design, its theoretical curve is given by $\ropt(\alpha)$; for the other three functions, we use the asymptotic predictions to be detailed in \sref{asymptotic} to evaluate their theoretical curves. Simulations results in the figure show the averages over 16 independent trials, with the error bars indicating $\pm 1$ standard deviation. The figure clearly demonstrates the improvement brought by the optimal design. In particular, we can see that the optimal preprocessing function \eref{T_Poiss} achieves the upper bound $\ropt(\alpha)$. Its performance dominates that of \eref{trimming}, \eref{subset}, and \eref{Tmm} uniformly over all $\alpha$.

\begin{figure}[t]
	\centering
	\includegraphics[width=0.7\linewidth]{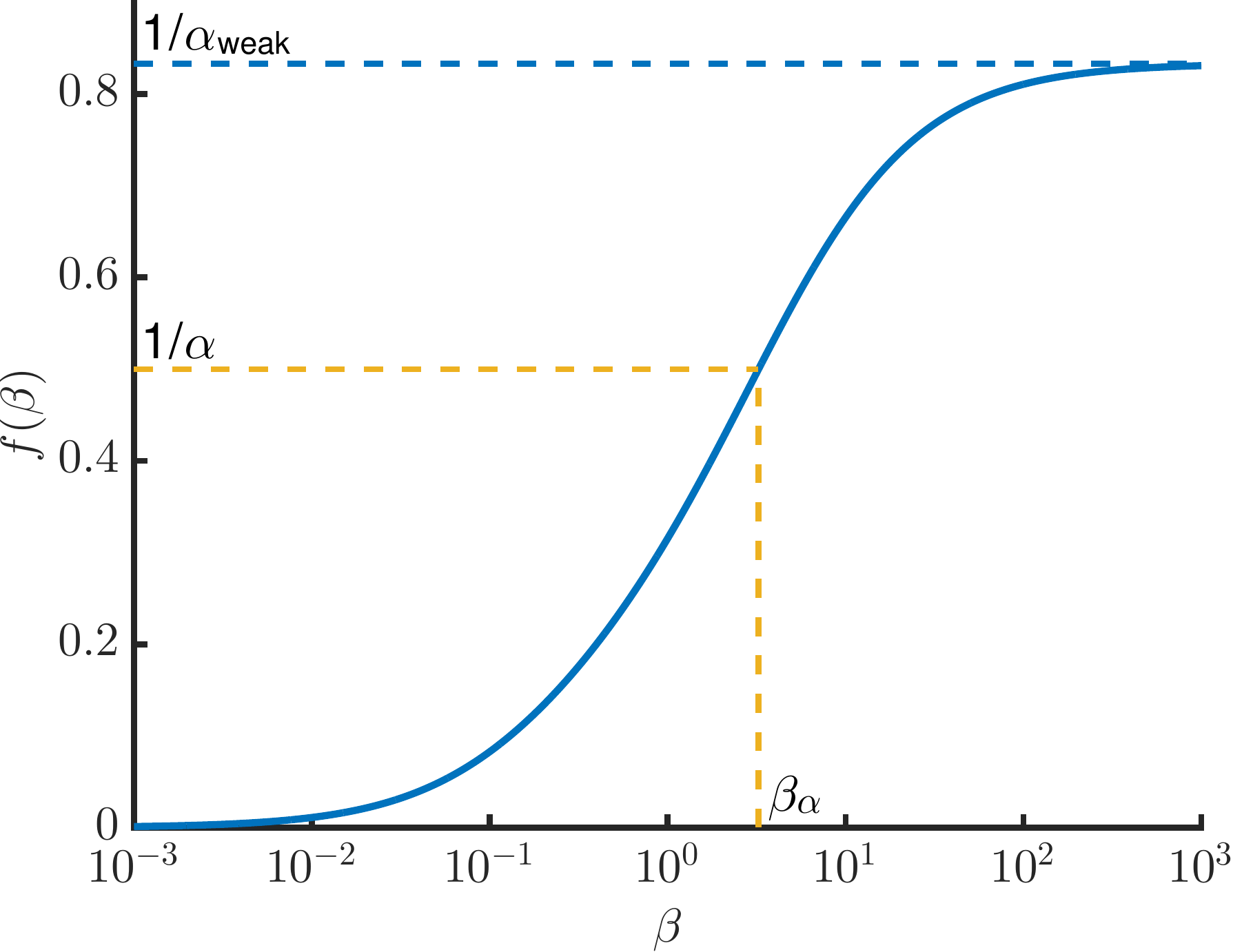}
	\caption{The function $f(\beta)$ in \eref{fbeta_Poisson} for $\kappa=5$. Here, $\lim_{\beta \to \infty}f(\beta)=1/\alpw=5/6$. For any  $\alpha>\alpw$, there is a unique solution $\beta_\alpha > 0$ to the equation $f(\beta) = 1/\alpha$.}
	\label{fig:fbeta} 
\end{figure}

\begin{figure}[t] 
	\centering
	\includegraphics[width=0.75\linewidth]{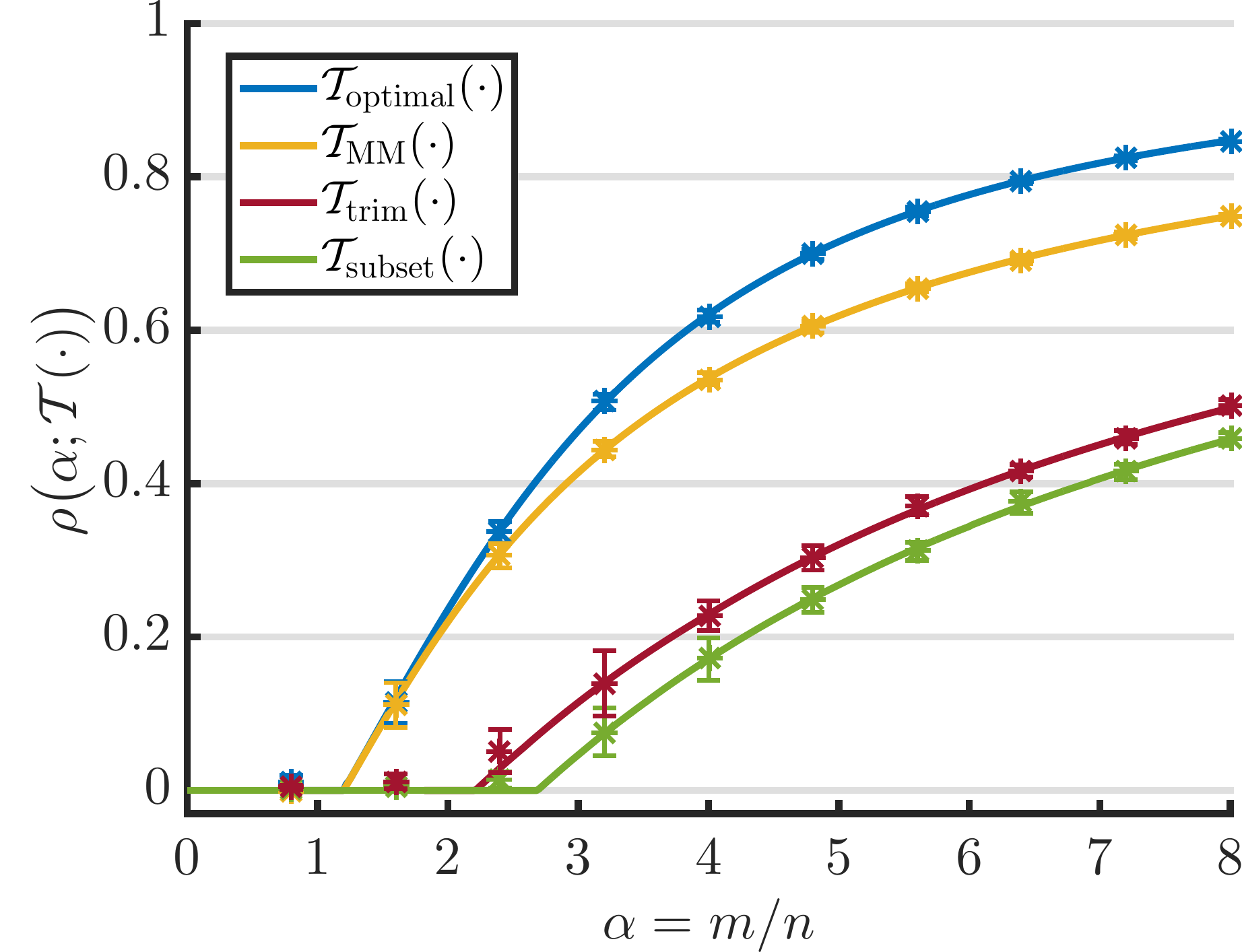}
	\caption{Analytical predictions and numerical simulations for the Poisson channel with different preprocessing functions. Numerical results are averaged over 16 independent trials.}
	\label{fig:Poisson}
\end{figure}

\begin{example}[The Gaussian channel] In the second example, we consider a sensing model with Gaussian noise:
\[
y_i = \max\big\{ \abs{\inprod{\va_i, \vxi}}^2+w_i, \text{ }0\big\},
\]
where $w_i\sim_\text{i.i.d.} \mathcal{N}(0,\sigma^2)$. When $\sigma^2>0$, we have
\[
\eta(y)= \begin{cases}  
\exp\big( \frac{\sigma^2}{2}-y \big) \Phi \big(\frac{y}{\sigma}-\sigma \big) ,&\text{if } y> 0, \\
\eta_0\,\delta(y), &\text{if } y=0,
\end{cases}
\]
and
\[
\mu(y)=\begin{cases}
(y-\sigma^2)\eta(y)+\frac{\sigma}{\sqrt{2\pi}} \exp\big(-\frac{y^2}{2\sigma^2}\big),&\text{if } y> 0, \\
\mu_0 \, \delta(y), &\text{if } y=0,
\end{cases}
\]
where $\Phi(\cdot)$ is the CDF of the standard normal distribution, $\delta(\cdot)$ denotes the Dirac delta function, and $\eta_0, \mu_0$ are two numerical constants defined as
\begin{equation}\label{eq:eta0}
\eta_0\bydef\int_{0}^\infty \Phi \big(-z/{\sigma}) e^{-z}\dif z = \frac{1}{2}-\exp \big(\frac{\sigma^2}{2}\big) \Phi(-\sigma)
\end{equation}
and
\begin{align}
\mu_0&\bydef \int_{0}^\infty \Phi \big(-z/{\sigma}) z e^{-z}\dif z  \label{eq:mu0}\\
&= \frac{1}{2}+(\sigma^2-1)\exp \big(\frac{\sigma^2}{2}\big) \Phi(-\sigma)-\frac{\sigma}{\sqrt{2\pi}}\nonumber
\end{align}
respectively. In Appendix~\ref{appendix:inf_Gaussian}, we show that
\begin{equation}\label{eq:inf_Gaussian}
\inf_{y \ge 0} \frac{\mu(y)}{\eta(y)}>0.
\end{equation}
It then follows from Theorem~\ref{thm:optimal} that the optimal performance curve $\rhoaT$ can be achieved by the following uniformly optimal preprocessing function
\begin{equation}\label{eq:T_gau}
\Topty =\begin{cases}
1-\bigg[y-\sigma^2+ \frac{\sigma \exp\big( -\frac{(y-\sigma^2)^2 }{2\sigma^2}\big)}{\sqrt{2\pi} \Phi\big(\frac{y}{\sigma}-\sigma\big)}\bigg]^{-1}, &\text{if }y>0\\
1-\frac{\eta_0}{\mu_0}, &\text{if }y=0.
\end{cases}
\end{equation}

%The comparison of $\rho(\mathcal{T},\alpha)$ for preprocessing functions indicated in \eref{T_gau} \eref{trimming} \eref{subset} \eref{Tmm} and the theoretical upper bound $\ropt(\alpha)$ are shown in fig 2. In our experiment, $n=4096$, $\sigma^2=1$. As we can see from fig 2, our proposed optimal preprocessing function \eref{T_gau} performs uniformly better than \eref{trimming} \eref{subset} \eref{Tmm} and achieves the theoretical upper bound.

Next, we consider the noiseless case, \emph{i.e.}, $y_i =  \abs{\inprod{\va_i, \vxi}}^2$ with $\sigma^2 = 0$. Here, the two functions $\eta(y)$ and $\mu(y)$ take much simpler forms:
\[
\eta(y) = e^{-y}  \quad\text{and}\quad \mu(y)= y e^{-y},
\]
but the challenge arises from the fact that, in this case,
\begin{equation}\label{eq:inf_noiseless}
\inf_{y \ge 0} \, \frac{\mu(y)}{\eta(y)}=\frac{\mu(0)}{\eta(0)}=0.
\end{equation}
As stated in Theorem~\ref{thm:optimal}, under \eref{inf_noiseless}, the optimal performance curve $\rhoaT$ cannot be achieved by any function $\T \in \mathcal{F}$. Later, in \sref{truncation}, we construct a family of preprocessing functions whose performance can arbitrarily approach $\rhoaT$. For any $\alpha > \alpw$, these functions take the form of 
\begin{equation}\label{eq:T_epsi}
\Taey=\frac{\caey}{1+\caey},
\end{equation}
where $0 < \varepsilon < 1$ is a parameter, 
\[
\caey=\max\left\{\vae \frac{\beta_\alpha (y-1)}{\beta_\alpha + y}, -1+\varepsilon \right\},
\]
and $\vae$ is a positive constant that can be uniquely determined by the following equation:
\[
\int_0^\infty \caey (y-1) e^{-y} \dif y={1}/{\alpha}.
\]
We show in \sref{truncation} that, as $\varepsilon \to 0$, the performance of $\Tae$ will converge to the optimal performance curve $\ropt(\alpha)$. This is demonstrated in \fref{noiseless}, where we compare the performance curves of the preprocessing function \eref{T_epsi} for three different values of $\varepsilon$ against the theoretical upper bound $\ropt(\alpha)$. We see that, when $\varepsilon=0.3$, the performance curves are already very close.

\begin{remark}
In practice, there is a trade-off between performance and computational cost when selecting the parameter $\varepsilon$. While a smaller $\varepsilon$ close to $0$ leads to better performance, it will also increase the magnitude of the negative eigenvalues of the data matrix $\mD$ in \eref{D_mtx}. The latter would slow down iterative algorithms such as power iterations that are often used to find the leading eigenvector of $\mD$. 
\end{remark}
%
%
%Another practical choice of preprocessing function is a simple truncation 
%\begin{equation}\label{eq:T_trunc}
%\mathcal{T}(y)=\max\Big\{ 1-\frac{\eta(y)}{\mu(y)}, c\Big\}=\max\Big\{ 1-\frac{1}{y}, c\Big\}.
%\end{equation}
%where $c$ is a tuning parameter usually set to be negative. There is the same trade-off between the performance and the computational cost when tuning parameter $c$.
\end{example}

\begin{figure}[t]
   \centering
   \includegraphics[width= 0.75\linewidth]{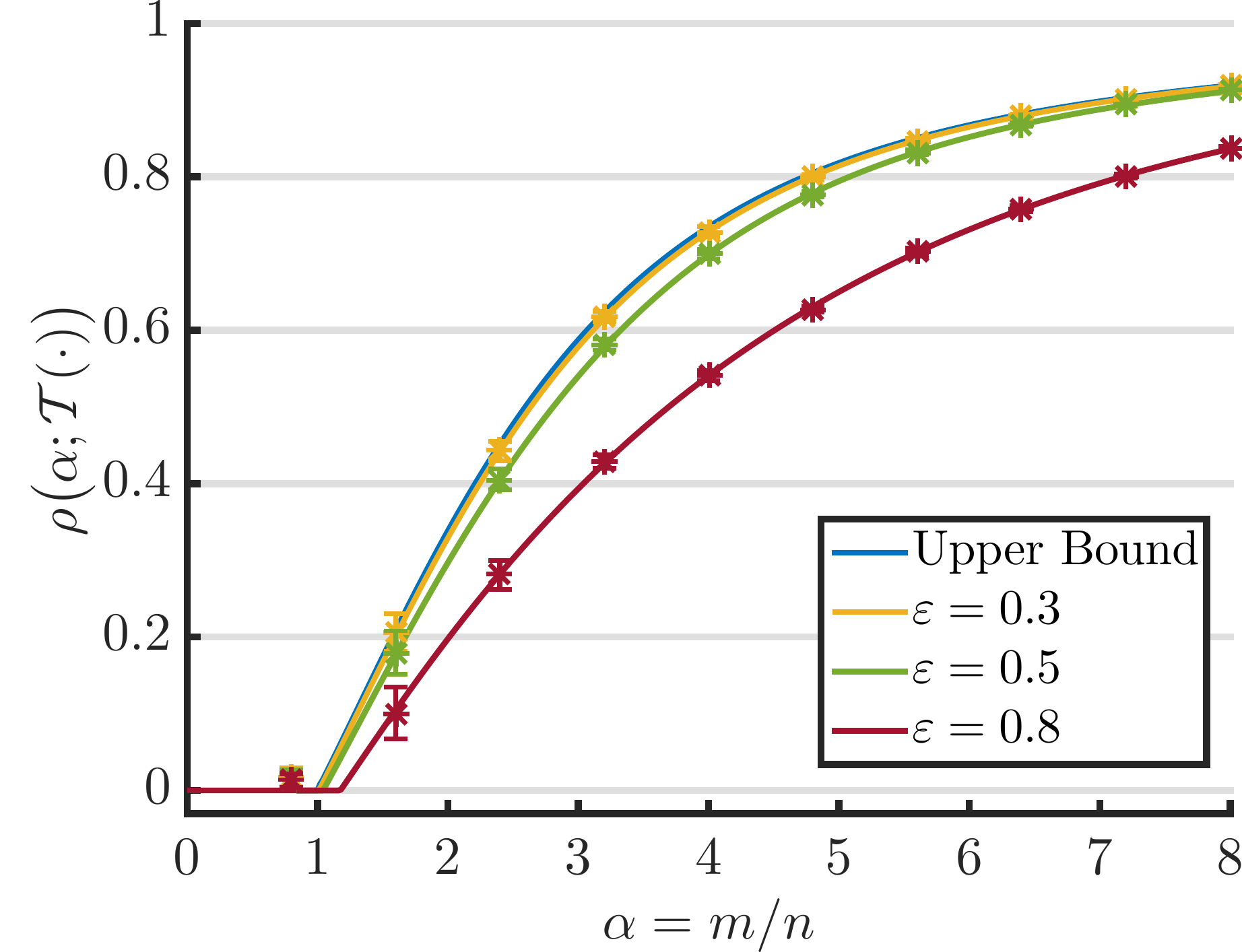}
   \caption{Analytical predictions and numerical simulations for the noiseless observation model. The blue curve corresponds to the theoretical upper bound that no preprocessing function in the feasibility set $\mathcal{F}$ can achieve. This upper bound can be approached by a family of preprocessing functions $\Tae$, as $\varepsilon \to 0$.}
   \label{fig:noiseless}
\end{figure}

\end{example}

%!TEX root = optimal_spectral.tex

\section{Technical Background}
\label{sec:technical}

\subsection{Asymptotic Characterizations of the Spectral Method}
\label{sec:asymptotic}

To set the stage for proving our main results on optimal design, we first review the precise asymptotic characterizations of the spectral method obtained in previous work \cite{LuL:18, MondelliM:17}.

Let $S$ and $Y$ be the random variables defined in \eref{S_rv} and \eref{Y_rv}, respectively. Recall the feasibility set $\mathcal{F}$ defined in \eref{feasible}. For any preprocessing function $\T \in \mathcal{F}$, the support of the probability measure of the random variable $\mathcal{T}(Y)$ is bounded. Let $\tau$ denote the upper boundary of the support, \emph{i.e.},
\[
\tau \bydef \sup_{y \in \Sy} \Ty < \infty.
\]
We consider two functions
\begin{equation}\label{eq:muf}
\muf(\lambda) \bydef \lambda \, \mathbb{E}_{S, Y}\bigg[\frac{ \TY \, \abs{S}^2}{\lambda - \TY}\bigg]
\end{equation}
and
\begin{equation}\label{eq:uf}
\uf_\alpha(\lambda) \bydef \lambda/\alpha  + \lambda \, \mathbb{E}_Y \bigg[\frac{\TY}{\lambda-\TY}\bigg],
\end{equation}
both defined on the open interval $(\ub, \infty)$. Within their domains, it is easy to check that both functions are convex and that $\muf(\lambda)$ is strictly decreasing. Consequently, if the following conditions
\begin{equation}\label{eq:fix_point}
\uf_\alpha(\lambda^\ast) = \muf(\lambda^\ast)
\end{equation}
and
\begin{equation}\label{eq:pos}
\uf'(\lambda^\ast) > 0,
\end{equation}
hold for some $\lambda^\ast \in (\tau, \infty)$, then that $\lambda^\ast$ must be unique.

\begin{theorem}[Asymptotic characterization \cite{LuL:18,MondelliM:17}]\label{thm:cos2}
Let the target signal $\vxi$ be an arbitrary vector in $\C^n$ with $\norm{\vxi} = 1$. Assume that the preprocessing function $\T \in \mathcal{F}$, and that the sensing vectors $\va_i \sim_\text{i.i.d.} \mathcal{CN}(\vec{0}, \mI_n)$. As $m, n \to \infty$ with $m/n \to \alpha \in (0, \infty)$, we have
\[
\frac{\abs{\inprod{\vxi, \vx_1}}^2}{\norm{\vxi}^2\norm{\vx_1}^2} \xrightarrow{\mathcal{P}} \rhoaT, 
\]
where the limit value on the right-hand side is
\begin{equation}\label{eq:limit_value}
\rhoaT \bydef \begin{cases}
\frac{\uf'_\alpha(\lm^\ast)}{\uf'_\alpha(\lm^\ast) - \muf'(\lm^\ast)}, &\text{if } \eref{fix_point}\eref{pos} \text{ hold for } \lambda^\ast > \tau\\
0, &\text{otherwise}.
\end{cases}
\end{equation}
\end{theorem}

%In particular, $\uf_\alpha(\lambda)$ achieves its minimum at a unique point denoted by 
%\begin{equation}\label{eq:barlm}
%\blm_\alpha \bydef \underset{\lambda > \tau}{\arg\,\min} \ \uf_\alpha(\lambda).
%\end{equation}
%Finally, let 
%\begin{equation}\label{eq:lf}
%\lf_\alpha(\lambda) \bydef \uf_\alpha\big(\lambda \vee \blm_\alpha\big)
%\end{equation}
%be a modification of $\uf_\alpha(\lambda)$. This new function is again defined for $\lambda \in (\ub, \infty)$.

The above theorem shows that, in the high-dimensional limit, the squared correlation between $\vxi$ and the estimate $\vx_1$  converges in probability to a deterministic value $\rhoaT$, which can be exactly computed as in \eref{limit_value}. Moreover, this asymptotic prediction exhibits a phase transition phenomenon: $\rhoaT$ is nonzero \emph{if and only if} \eref{fix_point} and \eref{pos} hold for some $\lambda^\ast > \tau$.

\begin{remark}
The asymptotic prediction stated in Theorem~\ref{thm:cos2} was first obtained in \cite{LuL:18} for the real-valued case. In that setting, the random variable $S$ in \eref{muf} and \eref{uf} should be drawn from $\mathcal{N}(0, 1)$. Additionally, \cite{LuL:18} makes an assumption that $\Ty \ge 0$. Later, Mondelli and Montanari showed that the same characterization holds for the complex-valued case. They also removed the restriction that $\Ty$ be positive by generalizing a result for spiked random positive semidefinite matrices \cite{BaiY:12} to spiked Hermitian matrices. In \cite{LuL:18, MondelliM:17}, there is also a technical assumption that, as $\lambda$ approaches $\tau$ from the right,
\begin{equation}\label{eq:ub_infty}
\lim_{\lambda \rightarrow \tau^+} \mathbb{E}\bigg[\frac{\TY}{(\lambda - \TY)^2}\bigg] = \lim_{\lambda \rightarrow \tau^+} \mathbb{E}\bigg[\frac{\TY \abs{S}^2}{\lambda - \TY}\bigg] = \infty.
\end{equation}
However, a close inspection of the arguments in \cite{LuL:18} (especially those in Propositions 2, 3, and 4 there) will show that the above assumption is unnecessary, if we present the limit value $\rhoaT$ in the form of \eref{limit_value}.
\end{remark}

\subsection{Reformulations of the Optimal Design Problem}
\label{sec:reform}

The optimal design problem formulated in \eref{opt_design_prob} seeks to find the supremum of  $\rhoaT$ over all preprocessing functions in the feasible set $\mathcal{F}$. Using the asymptotic characterizations of $\rhoaT$ given in \eref{limit_value}, we can convert the problem to
\[
\begin{aligned}
\sup_{\T \in \mathcal{F}} & \ \frac{\uf'_\alpha(\lm^\ast)}{\uf'_\alpha(\lm^\ast) - \muf'(\lm^\ast)}\\
\text{s.t.} &  \ \uf_\alpha(\lm^\ast) = \muf(\lm^\ast) \text{ and } \uf'_\alpha(\lm^\ast) > 0 \text{ for some } \lm^\ast > \tau.
\end{aligned}.
\]
For a given $\T$ and a given $\alpha$, it is possible that there is no $\lm^\ast > \tau$ satisfying the equality and inequality constraints in the above optimization problem. In that case, the value of the objective function is understood to be equal to $0$. 

Finding the supremum of $\frac{\uf'_\alpha(\lm^\ast)}{\uf'_\alpha(\lm^\ast) - \muf'(\lm^\ast)}$ is equivalent to finding the infimum of $\frac{- \muf'(\lm^\ast)}{\uf'_\alpha(\lm^\ast)}$, where the numerator $- \muf'(\lm^\ast)$ is positive due to the monotonicity of $\muf(\lm)$. We also rewrite the expectations in \eref{muf} and \eref{uf} in terms of $\eta(y)$ and $\mu(y)$ defined in \eref{eta} and \eref{mu}, respectively. Taking derivatives then gives us
\begin{align}
\inf_{\T \in \mathcal{F}} & \ \frac{\int_{\Sy}\big[\frac{\mathcal{T}(y)}{\lm^\ast-\mathcal{T}(y)}\big]^2\mu(y)\dif y}{1/\alpha-\int_{\Sy}[\frac{\mathcal{T}(y)}{\lm^\ast-\mathcal{T}(y)}]^2\eta(y)\dif y}\label{eq:opt_problem}\\
\text{s.t.} & \ \int_{\Sy} \tfrac{\mathcal{T}(y)}{\lm^\ast-\mathcal{T}(y)}[\mu(y)-\eta(y)]\dif y=1/\alpha \nonumber\\
\text{and}&\  \int_{\Sy}\big[\tfrac{\mathcal{T}(y)}{\lm^\ast-\mathcal{T}(y)}\big]^2\eta(y)\dif y <1/\alpha \text{ for some } \lambda^\ast > \tau.\nonumber
\end{align}

The problem in \eref{opt_problem} still appears unwieldy. To further simplify it, we observe that the objective function of \eref{opt_problem} as well as the equality and inequality constraints are all scale invariant since they are related to $\mathcal{T}(y)$ only through the ratio $ \frac{\mathcal{T}(y)}{\lm^\ast-\mathcal{T}(y)}$. Thus, if $(\Ty, \lambda^\ast)$ is a feasible solution satisfying the constraints, so will be $(a \Ty, a \lambda^\ast)$ for any constant $a > 0$. Meanwhile, the value of the objective function remains unchanged. Exploiting this invariance, we can always assume $\lambda^\ast = 1$, without loss of generality. Introducing a change of variables 
\begin{equation}\label{eq:cy}
c(y)\bydef \frac{\mathcal{T}(y)}{1-\mathcal{T}(y)},
\end{equation}
we can now simplify \eref{opt_problem} as
\begin{align}
(\mathrm{P_0})\quad V_0 \bydef \inf_{c(\cdot) \in \mathcal{F}_c} &\ \frac{\int_{\Sy}[c(y)]^2\mu(y)\dif y}{1/\alpha-\int_{\Sy}[c(y)]^2\eta(y)\dif y}\label{eq:org_opt0}\\
\text{s.t.} &\ \int_{\Sy} c(y)[\mu(y)-\eta(y)]\dif y=1/\alpha \label{eq:linear_const}\\
\text{and} &\ \int_{\Sy} [c(y)]^2\eta(y)\dif y <1/\alpha\label{eq:quad_const},
%& \text{ and } c(y)>-1
\end{align}
where $\mathcal{F}_c$ denotes the feasible set for $c(y)$, defined as
\[
\mathcal{F}_c \bydef \Big\{c(\cdot): 0 < \sup_{y \in \Sy} c(y) < \infty \text{ and } \inf_{y \in \Sy} c(y) > -1\Big\}.
\]
Note that the mapping \eref{cy}, or equivalently, $\Ty = \frac{c(y)}{1+c(y)}$, provides a one-to-one correspondance between $\mathcal{F}_c$ and
\[
\mathcal{F} \cap \set{\T: \sup_{y \in \Sy} \Ty < 1}.
\]
The additional constraint that $\sup_{y \in \Sy} \Ty < 1$ is both necessary and sufficient for our purpose, as we have fixed $\lambda^\ast = 1$. Moreover, we adopt the following notational convention: if there is no feasible $c(y) \in \mathcal{F}_c$ satisfying the constraints \eref{linear_const} and \eref{quad_const}, the value of the objective function $V_0 = +\infty$.

It will be more convenient to study
\begin{align}
(\mathrm{P_1})\quad V_1 \bydef \inf_{c(\cdot) \in \mathcal{H}} &\ \frac{\int_{\Sy}[c(y)]^2\mu(y)\dif y}{1/\alpha-\int_{\Sy}[c(y)]^2\eta(y)\dif y}\label{eq:org_opt1}\\
\text{s.t.} &\ \eref{linear_const} \text{ and } \eref{quad_const} \text{ hold},
\end{align}
where we simply relax $\mathcal{F}_c$ to a larger set 
\[
\mathcal{H} \bydef \Big\{c(\cdot): \int_{\Sy} [c(y)]^2 (\eta(y) + \mu(y)) \dif y < \infty\Big\}.
\]
We note that any function in $\mathcal{F}_c$ is finitely bounded. It then follows from \eref{unit_int} that the function must belong to $\mathcal{H}$. In the next section, we will present a closed-form solution to $(\mathrm{P_1})$. It forms the foundation of our proof of Theorem~\ref{thm:optimal}.

%!TEX root = optimal_spectral.tex

\section{Proof of Theorem~\ref{thm:optimal}}
\label{sec:proof}

In this section, we prove Theorem~\ref{thm:optimal} in three steps. First, we show in \sref{bound} that the right-hand side of \eref{rho_opt_beta} is an upper bound for the optimal performance curve $\ropt(\alpha)$. To establish equality, we consider two cases, depending on the value of $\inf_y \mu(y)/\eta(y)$. When $\inf_y \mu(y)/\eta(y) > 0$, we show in \sref{uniform} that the aforementioned upper bound can be achieved by the uniformly optimal solution given in \eref{T_uniform}, and that this optimal solution belongs to the feasible set $\mathcal{F}$. The remaining case, when $\inf_y \mu(y)/\eta(y) = 0$, is considered in \sref{truncation}, where we construct a family of functions $\Tae \in \mathcal{F}$ and show that their performance curves approach the upper bound.

\subsection{An Upper Bound for $\ropt(\alpha)$}
\label{sec:bound}
Following the discussions in \sref{reform}, we know that
\begin{equation}\label{eq:ropt_v0v1}
\ropt(\alpha) = \frac{1}{1 + V_0} \le \frac{1}{1 + V_1},
\end{equation}
where $V_0$ and $V_1$ are the optimal values of $(\mathrm{P_0})$ and $(\mathrm{P_1})$, respectively. That $V_1 \le V_0$ is due to the fact that the set $\mathcal{F}_c$ in $(\mathrm{P_0})$ is a subset of $\mathcal{H}$ in $(\mathrm{P_1})$. In what follows, we present a solution of $(\mathrm{P_1})$.

For each $\beta > 0$, the (sublevel set) condition
\[
\frac{\int_{\Sy}[c(y)]^2\mu(y)\dif y}{1/\alpha-\int_{\Sy}[c(y)]^2\eta(y)\dif y} \le \beta
\]
is equivalent to $\int_{\Sy}[c(y)]^2[\mu(y)/\beta+\eta(y)]\dif y \le 1/\alpha$. To lighten the notation, we define two functionals
\begin{equation}\label{eq:Qc}
\Qc \bydef \int_{\Sy}[c(y)]^2[\mu(y)/\beta+\eta(y)]\dif y,
\end{equation}
and
\begin{equation}\label{eq:Lc}
\Lc \bydef \int_{\Sy} c(y)[\mu(y)-\eta(y)]\dif y.
\end{equation}
It is easy to see that $(\mathrm{P_1})$ is equivalent to
\begin{equation}\label{eq:set_opt2}
V_1 = \inf_{\Omega_\beta \text{ is nonempty} } \beta,
\end{equation}
where
\begin{equation}\label{eq:omega_beta}
\Omega_\beta \bydef  \big\{ c(\cdot) \in \mathcal{H}: \Qc \leq \tfrac{1}{\alpha} \text{ and }  \Lc=\tfrac{1}{\alpha} \big\}.
\end{equation}
Note that we can omit the quadratic constraint in \eref{quad_const} as it is implied by $\Qc \leq \tfrac{1}{\alpha}$.

\begin{lemma}\label{lemma:opt1}
For any $\beta > 0$,
\begin{equation}\label{eq:optl1}
\begin{aligned}
\frac{1}{\alpha^2 f(\beta)} = \min_{c(\cdot ) \in \mathcal{H}} &\text{  }\Qc\\
\text{s.t.} &\text{  }\Lc= 1/\alpha
\end{aligned},
\end{equation}
where $f(\beta)$ is the function defined in \eref{fbeta}. Moreover, the optimal solution is given by
\begin{equation}\label{eq:c_mn}
c^\ast(y) = \frac{\mu(y) - \eta(y)}{[\alpha f(\beta)](\eta(y) + \mu(y)/\beta)}.
\end{equation}
\end{lemma}
\begin{IEEEproof}
Since $\eta(y) + \mu(y)/\beta$ is a nonnegative function, we can define a \emph{weighted} $L^2$ function space, where the inner product between two functions $f_1(y)$ and $f_2(y)$ is
\[
\inprod{ f_1(y),f_2(y)}_\beta \bydef \int_{\Sy} f_1(y) f_2(y) [\eta(y) + \mu(y)/\beta] \dif y.
\]
The optimization problem \eref{optl1} then becomes that of finding a minimum norm solution on a linear variety, \emph{i.e.},
\[
\begin{aligned}
\min_{c(\cdot)} & \text{  } \inprod{c(y), c(y)}_\beta \\
\text{s.t.} &  \text{  } \inprod{c(y), \frac{ \mu(y) - \eta(y)}{\eta(y) + \mu(y)/\beta} }_\beta = 1/\alpha
\end{aligned}.
\]
The optimal solution should take the form of 
\[
c^\ast(y)=v \frac{ \mu(y) - \eta(y)}{\eta(y) + \mu(y)/\beta},
\]
where the scaling constant $v$ is determined by the constraint $\int_{\Sy} c(y) [ \mu(y) - \eta(y)] \dif y = 1/\alpha$. 
Solving this equation gives $v=1/[\alpha f(\beta)]$ and thus \eref{c_mn}, the squared norm of which gives us the left-hand side of \eref{optl1}.
\end{IEEEproof}

Applying Lemma~\ref{lemma:opt1}, we know that the set $\Omega_\beta$ in \eref{omega_beta} is nonempty if and only if
\begin{equation}\label{eq:f_beta_alpha}
f(\beta) \ge 1/\alpha.
\end{equation}
From its definition given in \eref{fbeta}, $f(\beta)$ is a strictly increasing function, with
\[
\lim_{\beta \to 0} f(\beta) = 0 \quad\text{and}\quad \lim_{\beta \to \infty} f(\beta) = 1/\alpw.
\]
 
Consequently, for $\alpha \le \alpha_{\text{weak}}$, the condition \eref{f_beta_alpha} cannot be satisfied by any finite $\beta$. In this case, the optimal value of \eref{set_opt2} is $V_1 = \infty$. Using \eref{ropt_v0v1}, we conclude that
\[
\ropt(\alpha) = 0 \quad \text{ for } \alpha \le \alpw,
\]
recovering the results derived in \cite{MondelliM:17}.

When $\alpha>\alpha_{\text{weak}}$, there is a unique $\beta_\alpha > 0$ such that $f(\beta_\alpha) = 1/\alpha$. It follows from the monotonicity of $f(\beta)$ that $V_1 = \beta_\alpha$. Substituting this into \eref{ropt_v0v1}, we get
\begin{equation}\label{eq:upper_bnd}
\rho_\text{optimal}(\alpha) \le (1 + \beta_\alpha)^{-1} \quad \text{for } \alpha > \alpha_\text{weak}.
\end{equation}
In the next two subsections, we show that this inequality is in fact an equality.

\subsection{Uniformly Optimal Preprocessing Function}
\label{sec:uniform}

We first consider the case when $\inf_y \mu(y)/\eta(y)>0$. For each $\alpha > \alpw$, we know from Lemma~\ref{lemma:opt1} that the upper bound in \eref{upper_bnd} is achieved by 
\begin{equation}\label{eq:c_ast_p1}
c^\ast(y) = \frac{\mu(y) - \eta(y)}{[\alpha f(\beta_\alpha)](\eta(y) + \mu(y)/\beta_\alpha)} = \frac{\mu(y) - \eta(y)}{\eta(y) + \mu(y)/\beta_\alpha},
\end{equation}
where the second equality is due to the fact that $f(\beta_\alpha) = 1/\alpha$. It is easy to verify that
\begin{equation}\label{eq:c_ast_bnd}
-1 \le \inf_{y \in \Sy} c^\ast(y) \le \sup_{y \in \Sy} c^\ast(y) \le \beta_\alpha.
\end{equation}
From \eref{cy}, the corresponding preprocessing function is
\[
\mathcal{T}(y)=\frac{c^\ast(y)}{1+c^\ast(y)}=\frac{1}{1+1/\beta_\alpha}\big[1-\frac{\eta(y)}{\mu(y)}\big].
\]
As mentioned in \sref{reform}, the performance of the spectral algorithm is scale invariant. So a scaled version
\begin{equation}\label{eq:topty1}
\Topty=1-\frac{\eta(y)}{\mu(y)}
\end{equation}
can achieve the same performance. Next, we show that $\Topty \in \mathcal{F}$. This would then imply that $\ropt(\alpha) \ge (1+\beta_\alpha)^{-1}$, which, together with \eref{upper_bnd}, gives us \eref{rho_opt_beta}.

Since $\eta(y)\geq 0$ and $\mu(y)\geq 0$, 
\[
\tau = \sup_{y \in \Sy} \Topty \le 1.
\]
Under the assumption that $\inf_y \mu(y)/\eta(y)>0$, we also have $\inf_{y \in \Sy} \Topty > -\infty$. What remains to be shown is that $\tau > 0$. To that end, we first note that $\eta(y)$ cannot be identically equal to $\mu(y)$, as otherwise the weak reconstruction threshold $\alpw = \infty$. Meanwhile, \eref{unit_int} implies that
\[
\int_{\Sy} (\eta(y) - \mu(y)) \dif y = 0.
\]
Thus, there must exist $y$ for which $\mu(y)>\eta(y)$. This then guarantees that $\tau > 0$.

\subsection{Truncated Preprocessing Functions}
\label{sec:truncation}

In this section, we consider the case when
\begin{equation}\label{eq:infy0}
\inf_y \mu(y)/\eta(y)=0.
\end{equation}
In this case, the function $\Topt$ in \eref{topty1} is not lower bounded, and thus it is not in the feasibility set $\mathcal{F}$. This then implies that the optimal performance curve $\ropt(\alpha)$ in \eref{rho_opt_beta} cannot be achieved by any function in $\mathcal{F}$. 

To see this point, we suppose that there exists some function $\widetilde{c}(y) \in \mathcal{F}_c$ that achieves the infimum $V_0$ in \eref{org_opt0}. Since $V_0$ is equal to the infimum $V_1$ in \eref{org_opt1} and since $\mathcal{F}_c \subset \mathcal{H}$, this would mean that $\widetilde{c}(y)$ is also an optimal solution of $(\mathrm{P_1})$. We now have a contradiction: $(\mathrm{P_1})$ admits a unique optimal solution $c^\ast(y)$ given by \eref{c_ast_p1} which does not belong to $\mathcal{F}_c$ under \eref{infy0}, and thus $c^\ast(\cdot) \neq \widetilde{c}(\cdot)$.

In what follows, we show that the family of preprocessing functions $\set{\Tae \in \mathcal{F}}_{0<\varepsilon<1}$ defined in \eref{c_ast_e} can approach the optimal performance curve, \emph{i.e.},
\begin{equation}\label{eq:conv_eps}
\lim_{\varepsilon \to 0} \,\rho(\alpha;\Tae) = \ropt(\alpha)
\end{equation}
for all $\alpha > \alpw$.

We start by showing that the scaler $v_\alpha^\varepsilon$ in \eref{c_ast_e} can indeed be uniquely determined by the linear constraint $\mathcal{L}(c_\alpha^\epsilon(\cdot)) = 1/\alpha$. To that end, we define
\begin{equation}\label{eq:cvy}
c_v(y) = \max\left\{v \frac{\mu(y) - \eta(y)}{\eta(y) + \mu(y)/\beta_\alpha}, -1+\varepsilon \right\},
\end{equation}
with $v > 0$ being a varying parameter, and examine
\begin{align}
&\Lcv \bydef \int_{\Sy}c_v(y)[\mu(y)-\eta(y)]\dif y \label{eq:Lcv}\\
&=\int_{\Sy} \max\left\{v \frac{\mu(y) - \eta(y)}{\eta(y) + \mu(y)/\beta_\alpha}, -1+\varepsilon \right\}[\mu(y)-\eta(y)]\dif y.\nonumber
\end{align}
We note that $-1+\varepsilon<0$ and $c_v(y)[\mu(y)-\eta(y)]\geq 0$ for all $y\in \Sy$. A moment of thought will convince us that $\Lcv$ is an increasing function of $v$ for $v \ge 0$. Moreover, when $v = 0$, $\Lcv = 0$. To study the limit of the function as $v \to \infty$, we denote by $\Gamma_Y^+$ the subset over which $\mu(y) > \eta(y)$. We have
\[
\begin{aligned}
\Lcv &\geq \int_{\Gamma_Y^+} c_v(y) [\mu(y)-\eta(y)]\dif y \\
&=v \int_{\Gamma_Y^+} \frac{[\mu(y) - \eta(y)]^2}{\eta(y) + \mu(y)/\beta_\alpha} \dif y,
\end{aligned}
\]
which tends to $\infty$ as $v \to \infty$. It then follows that there exists a unique positive solution $v_\alpha^\varepsilon$ to the equation $\Lcv = 1/\alpha$.

To establish \eref{conv_eps}, we recall our reformulations of the optimal design problem presented in \sref{reform}. Given the equivalence of the optimal design problem and the optimization problem $(\mathrm{P_0})$ in \eref{org_opt0}, our tasks boil down to showing that (1) $c_\alpha^\varepsilon(y) \in \mathcal{F}_c$ and (2) for each $\alpha > \alpw$,
\begin{equation}\label{eq:fin1}
\lim_{\varepsilon \rightarrow 0} \int_{\Sy} [c_\alpha^\varepsilon(y)]^2\eta(y)\dif y= \int_{\Sy} [c^\ast(y)]^2\eta(y)\dif y,
\end{equation}
\begin{equation}\label{eq:fin2}
\lim_{\varepsilon \rightarrow 0} \int_{\Sy} [c_\alpha^\varepsilon(y)]^2\mu(y)\dif y= \int_{\Sy} [c^\ast(y)]^2\mu(y)\dif y,
\end{equation}
where $c^\ast(y)$ is the optimal solution given in \eref{c_ast_p1}. Note that the linear constraint \eref{linear_const} in $(\mathrm{P_0})$ is always satisfied, due to the way we set the scalar $v_\alpha^\varepsilon$ in \eref{c_ast_e}. The quadratic constraint \eref{quad_const} will also be satisfied for all sufficiently small $\varepsilon$, given the the convergence in \eref{fin2} and the fact that we have $\int_{\Sy} [c^\ast(y)]^2\eta(y)\dif y <1/\alpha$.

By its definition in \eref{c_ast_e}, it is easy to see that
\begin{equation}\label{eq:c_ae_bnd}
-1+\varepsilon \le \inf_{y \in \Sy} c_\alpha^\varepsilon(y) \le \sup_{y \in \Sy} c_\alpha^\varepsilon(y) \le v_\alpha^\varepsilon \beta_\alpha.
\end{equation}
Moreover, since $\mu(y) > \eta(y)$ over a nonempty subset of $\Sy$, we have $\sup_{y \in \Sy} c_\alpha^\varepsilon(y) > 0$. Thus, we can verify that $c_\alpha^\varepsilon(y) \in \mathcal{F}_c$. 

Next, we show that $v_\alpha^\varepsilon \to 1$ as $\varepsilon \to 0$. Recall that
\begin{equation}\label{eq:linear_const_2}
\mathcal{L}(c_\alpha^\varepsilon(\cdot)) = \mathcal{L}(c^\ast(\cdot)) = 1/\alpha.
\end{equation}
The latter equality implies that 
\[
\int_{\Sy} \max\left\{c^\ast(y), -1+\varepsilon \right\}[\mu(y)-\eta(y)]\dif y \le 1/\alpha,
\]
which, using the notation introduced in \eref{cvy} and \eref{Lcv}, can be written as $\mathcal{L}(c_1(\cdot)) \le 1/\alpha$. Since $\Lcv$ is an increasing function of $v$, we must have $v_\alpha^\varepsilon \ge 1$. 

Define two sets $\Gamma_1$ and $\Gamma_2$, with $\Gamma_1\cup \Gamma_2 =\Sy$, such that $c_\alpha^\varepsilon(y)=-1+\varepsilon$ for $y\in \Gamma_1$ and $ c_\alpha^\varepsilon(y)=v_\alpha^\varepsilon \frac{\mu(y) - \eta(y)}{\eta(y) + \mu(y)/\beta_\alpha}$ for $y\in \Gamma_2$. Similarly, define a subset $\Gamma'_1$ such that $ c^\ast(y)\leq -1+\varepsilon$ for $y\in \Gamma'_1$. We can easily verify that $\Gamma'_1 \subset \Gamma_1$, since $c_\alpha^\varepsilon(y) = \max\set{v_\alpha^\varepsilon c^\ast(y), -1 + \varepsilon}$ and $v_\alpha^\varepsilon \ge 1$.  It follows from \eref{linear_const_2} that
\begin{equation}\label{eq:ineq}
\begin{aligned}
0=&\mathcal{L}(c^\ast(\cdot)) - \mathcal{L}(c_\alpha^\varepsilon(\cdot))\\
=&\int_{\Gamma'_1}\left[ c^\ast(y)-\left(-1+\varepsilon\right) \right]\left( \mu(y) - \eta(y) \right)\dif y \\
&+\int_{\Gamma_1\setminus \Gamma'_1}\left[c^\ast(y)-\left(-1+\varepsilon\right) \right] \left( \mu(y) - \eta(y) \right) \dif y  \\
&-(v_\alpha^\varepsilon-1) \int_{\Gamma_2}  \frac{\mu(y) - \eta(y)}{\eta(y) + \mu(y)/\beta_\alpha} \left( \mu(y) - \eta(y) \right) \dif y.
\end{aligned}
\end{equation}

For $y\in \Gamma_1\setminus \Gamma'_1 $, we know $\mu(y) - \eta(y)<0$ and  $c^\ast(y) \ge -1+\varepsilon$. Thus, 
\begin{equation}\label{eq:bnd_1}
\int_{\Gamma_1\setminus \Gamma'_1}\left[c^\ast(y) - (-1+\varepsilon) \right] \left( \mu(y) - \eta(y) \right) \dif y\le 0.
\end{equation}
Also for $y\in \Gamma'_1$, we have  $-1 \leq c^\ast(y) \le \left(-1+\varepsilon\right)$, so
\begin{align}
&\abs{\int_{\Gamma'_1}\left[ c^\ast(y)-\left(-1+\varepsilon\right) \right]\left( \mu(y) - \eta(y) \right)\dif y} \nonumber\\
& \leq \varepsilon \int_{\Sy} [\mu(y) + \eta(y)] \dif y = 2 \varepsilon.\label{eq:bnd_2}
\end{align}

Substituting \eref{bnd_1} and \eref{bnd_2} into \eref{ineq}, we get
\[
(v_\alpha^\varepsilon-1) \int_{\Gamma_2}  \frac{[\mu(y) - \eta(y)]^2}{\eta(y) + \mu(y)/\beta_\alpha} \dif y \leq 2\varepsilon.
\] 
Let $\Gamma_Y^+$ be the subset such that $\mu(y) > \eta(y)$ for $y \in \Gamma_Y^+$. We must have $\Gamma_Y^+ \subset \Gamma_2$. It follows that
\[
1 \le v_\alpha^\varepsilon \le 1 + 2\varepsilon \Big(\int_{\Gamma_Y^+}  \frac{[\mu(y) - \eta(y)]^2}{\eta(y) + \mu(y)/\beta_\alpha} \dif y\Big)^{-1}
\]
and thus $v_\alpha^\varepsilon \to 1$ as $\varepsilon \to 0$. 

What remain to be shown are \eref{fin1} and \eref{fin2}. The proofs for the two cases are essentially identical, so we focus on establishing \eref{fin1}, as follows:
\[
\begin{aligned}
&\abs{\int_{\Sy} ([c_\alpha^\varepsilon(y)]^2 - [c^\ast(y)]^2) \eta(y)\dif y}\\
&\le \int_{\Sy} \abs{c_\alpha^\varepsilon(y) - c^\ast(y)} \big(\abs{c_\alpha^\varepsilon(y)} + \abs{c^\ast(y)}\big)\eta(y)\dif y\\
&\overset{(a)}{\le} C_\varepsilon \int_{\Sy} \abs{c_\alpha^\varepsilon(y) - c^\ast(y)}\eta(y)\dif y\\
&\le C_\varepsilon \int_{\Sy} \abs{c_\alpha^\varepsilon(y) - \max\set{c^\ast(y), -1+\varepsilon}}\eta(y)\dif y\\
&\qquad\qquad + C_\varepsilon \int_{\Sy} \abs{\max\set{c^\ast(y), -1+\varepsilon} - c^\ast(y)}\eta(y)\dif y\\
&\overset{(b)}{\le} C_\varepsilon \abs{v_\alpha^\epsilon - 1} \int_{\Sy} \abs{c^\ast(y)} \eta(y) \dif y + C_\varepsilon \cdot \varepsilon \int_{\Sy} \eta(y) \dif y\\
&\overset{(c)}{\le} C_\varepsilon \abs{v_\alpha^\epsilon - 1} \beta_\alpha + C_\varepsilon \cdot \varepsilon.
\end{aligned}
\]
Here, to obtain (a), we have used the boundedness of $c_\alpha^\varepsilon(y)$ as given in \eref{c_ae_bnd} and that of $c^\ast(y)$ as given in \eref{c_ast_bnd}. Consequently, it is sufficient to set the constant to be
\[
C_\varepsilon = 2 + (v_\alpha^\varepsilon + 1) \beta_\alpha.
\]
To reach (b), we have used the following properties:
\[
\abs{\max\set{x, -1+\varepsilon} - \max\set{y, -1+\varepsilon}} \le \abs{x - y}
\]
and 
\[
\abs{\max\set{x, -1+\varepsilon} - x} \le \varepsilon
\]
for any $x, y \ge -1$. Finally, the inequality (c) follows from the boundedness of $\abs{c^\ast(y)}$ and the fact that $\int_{\Sy} \eta(y) \dif y = 1$. Since $v_\alpha^\varepsilon \to 1$ as $\varepsilon \to 0$, we have \eref{fin1}.

%\begin{remark}
%In most practical settings, we will not run into the problem of $\inf_y \frac{\mu(y)}{\eta(y)} = 0$. When $\inf_y \frac{\mu(y)}{\eta(y)} = 0$, the unboundedness of function $1-\frac{\eta(y)}{\mu(y)}$ not only breaks the analysis but also makes the spectral method computationally expensive because a very small eigenvalue can make algorithm like power iteration converge very slowly. 
%\end{remark}

%!TEX root = optimal_spectral.tex

\appendix

\section*{}

\subsection{A Lower Bound on $\alpw$}
\label{appendix:alpw}

In this appendix, we show that the integral on the right-hand side of \eref{alpw} is always well-defined. In particular, we establish the following fundamental lower bound on the weak reconstruction threshold.

\begin{proposition}
For any sensing model given in \eref{model}, we have
\[
\alpw \ge 1,
\]
where the lower bound is achieved when $y_i = \abs{\inprod{\va_i, \vxi}}^2$.
\end{proposition}
\begin{IEEEproof}
Using the definition in \eref{alpw}, we have
\[
\begin{aligned}
\alpha^{-1}_\mathrm{weak}&=\int_{\Sy}\frac{\big( \mu(y)-\eta(y)\big)^2}{\eta(y)}\dif y \\
&=\int_{\Sy} \frac{[\mu(y)]^2}{\eta(y)} \dif y -2\int_{\Sy}\mu(y)\dif y+ \int_{\Sy}\eta(y)\dif y \\
&=\int_{\Sy} \frac{[\mu(y)]^2}{\eta(y)} \dif y -1,
\end{aligned}
\]
where the last equality is due to \eref{unit_int}. Thus, we just need to show that $\int_{\Sy} \frac{[\mu(y)]^2}{\eta(y)} \dif y \leq 2$.

Let $Z = \abs{S}$, where $S \sim \mathcal{CN}(0, 1)$. Let $g(z)$ denote the density function of $Z$. We have
\[
\begin{aligned}
{[\mu(y)]}^2 &= \Big[\int_0^\infty z^2 \, p(y \mid  z)g(z) \dif z\Big]^2\\
&\le \Big[\int_0^\infty z^4 \, p(y \mid  z)g(z) \dif z\Big] \Big[\int_0^\infty  p(y \mid  z)g(z) \dif z\Big]\\
&= \Big[\int_0^\infty z^4 \, p(y \mid  z)g(z) \dif z\Big] \eta(y),
\end{aligned}
\]
where the bound is due to the Cauchy-Schwarz inequality. Using this upper bound, we have
\begin{align}
\int_{\Sy} \frac{{[\mu(y)]}^2}{\mu(y)} \dif y &\le \int_0^\infty z^4 g(z) \int_{\Sy} p(y \mid  z) \dif y \dif z \nonumber\\
	&=\int_0^\infty z^4 g(z) \dif z \nonumber\\
	&= \mathbb{E}[Z^4] = 2.\label{eq:alpw_bnd}
\end{align}
For the noiseless channel described in \sref{examples}, we have $\eta(y)=e^{-y}$ and $\mu(y)=ye^{-y}$ for $y\geq 0$. So
$\int_{\Sy} \frac{[\mu(y)]^2}{\eta(y)} \dif y = \int_0^\infty y^2e^{-y}=2$, which achieves the bound in \eref{alpw_bnd}.
\end{IEEEproof}

\subsection{Proof of \eref{inf_Gaussian}}
\label{appendix:inf_Gaussian}

We note that
\[
\inf_{y \ge 0} \frac{\mu(y)}{\eta(y)} = \min\set{\inf_{y > 0} \frac{\mu(y)}{\eta(y)}, \frac{\mu_0}{\eta_0}}.
\]
By construction, $\eta_0, \mu_0$ as defined in \eref{eta0} and \eref{mu0} are both positive. Thus, to show \eref{inf_Gaussian}, we just need to prove that $\inf_{y > 0} \frac{\mu(y)}{\eta(y)} > 0$. To that end, we note that, for $y > 0$,
\begin{equation}\label{eq:mu/eta}
\frac{\mu(y)}{\eta(y)}=\sigma h(y/\sigma - \sigma),
\end{equation}
where $h(x) \bydef x + \frac{\Phi'(x)}{\Phi(x)}$, with $\Phi(x)$ and $\Phi'(x)$ denoting the CDF and PDF of the standard normal distribution, respectively. The function $h(x)$ is related to the inverse Mill's ratio. It is a strictly increasing function, and $h(x) > 0$ for all $x$. See, \emph{e.g.}, \cite{sampford1953some} for a proof. It follows that
\[
\inf_{y > 0} \frac{\mu(y)}{\eta(y)} = \lim_{y \to 0^+} \frac{\mu(y)}{\eta(y)} = \sigma h(-\sigma) > 0.
\]

\bibliographystyle{IEEEtran}
\bibliography{refs}

\begin{thebibliography}{10}
\providecommand{\url}[1]{#1}
\csname url@samestyle\endcsname
\providecommand{\newblock}{\relax}
\providecommand{\bibinfo}[2]{#2}
\providecommand{\BIBentrySTDinterwordspacing}{\spaceskip=0pt\relax}
\providecommand{\BIBentryALTinterwordstretchfactor}{4}
\providecommand{\BIBentryALTinterwordspacing}{\spaceskip=\fontdimen2\font plus
\BIBentryALTinterwordstretchfactor\fontdimen3\font minus
  \fontdimen4\font\relax}
\providecommand{\BIBforeignlanguage}[2]{{%
\expandafter\ifx\csname l@#1\endcsname\relax
\typeout{** WARNING: IEEEtran.bst: No hyphenation pattern has been}%
\typeout{** loaded for the language `#1'. Using the pattern for}%
\typeout{** the default language instead.}%
\else
\language=\csname l@#1\endcsname
\fi
#2}}
\providecommand{\BIBdecl}{\relax}
\BIBdecl

\bibitem{Netrapalli:2013}
P.~Netrapalli, P.~Jain, and S.~Sanghavi, ``Phase retrieval using alternating
  minimization,'' in \emph{Advances in {Neural} {Information} {Processing}
  {Systems}}, 2013, pp. 2796--2804.

\bibitem{Candes:2015}
E.~J. Candes, X.~Li, and M.~Soltanolkotabi, ``Phase retrieval via {Wirtinger}
  flow: {Theory} and algorithms,'' \emph{Information Theory, IEEE Transactions
  on}, vol.~61, no.~4, pp. 1985--2007, 2015.

\bibitem{Chen:2015}
Y.~Chen and E.~J. Candes, ``Solving {Random} {Quadratic} {Systems} of
  {Equations} {Is} {Nearly} as {Easy} as {Solving} {Linear} {Systems},''
  \emph{arXiv preprint arXiv:1505.05114}, 2015.

\bibitem{LiGL:15}
G.~Li, Y.~Gu, and Y.~M. Lu, ``Phase retrieval using iterative projections:
  Dynamics in the large systems limit,'' in \emph{Proc. Allerton Conference on
  Communication, Control and Computing.}, Monticello, IL, Oct 2015.

\bibitem{zhang2016provable}
H.~Zhang, Y.~Chi, and Y.~Liang, ``Provable non-convex phase retrieval with
  outliers: Median truncated {W}irtinger flow,'' in \emph{International
  conference on machine learning}, 2016, pp. 1022--1031.

\bibitem{WangGY:2016}
G.~Wang, G.~B. Giannakis, and Y.~C. Eldar, ``Solving {Systems} of {Random}
  {Quadratic} {Equations} via {Truncated} {Amplitude} {Flow},''
  \emph{arXiv:1605.08285}, May 2016.

\bibitem{ChiL:16}
Y.~Chi and Y.~M. Lu, ``Kaczmarz method for solving quadratic equations,''
  \emph{IEEE Signal Process. Lett.}, vol.~23, no.~9, 2016.

\bibitem{ma2018optimization}
J.~Ma, J.~Xu, and A.~Maleki, ``Optimization-based amp for phase retrieval: The
  impact of initialization and $\ell\_2 $-regularization,'' \emph{arXiv
  preprint arXiv:1801.01170}, 2018.

\bibitem{Candes:2013xy}
E.~J. Candes, T.~Strohmer, and V.~Voroninski, ``Phaselift: {Exact} and stable
  signal recovery from magnitude measurements via convex programming,''
  \emph{Communications on Pure and Applied Mathematics}, vol.~66, no.~8, pp.
  1241--1274, 2013.

\bibitem{Candes:2014ty}
E.~J. Candes and X.~Li, ``Solving quadratic equations via {PhaseLift} when
  there are about as many equations as unknowns,'' \emph{Foundations of
  Computational Mathematics}, vol.~14, no.~5, pp. 1017--1026, 2014.

\bibitem{Jaganathan:2013zl}
K.~Jaganathan, S.~Oymak, and B.~Hassibi, ``Sparse phase retrieval: {Convex}
  algorithms and limitations,'' in \emph{Information {Theory} {Proceedings}
  ({ISIT}), 2013 {IEEE} {International} {Symposium} on}.\hskip 1em plus 0.5em
  minus 0.4em\relax IEEE, 2013, pp. 1022--1026.

\bibitem{Waldspurger:2015rz}
I.~Waldspurger, A.~d'Aspremont, and S.~Mallat, ``Phase recovery, maxcut and
  complex semidefinite programming,'' \emph{Mathematical Programming}, vol.
  149, no. 1-2, pp. 47--81, 2015.

\bibitem{Li:92}
K.-C. Li, ``On principal hessian directions for data visualization and
  dimension reduction: Another application of {Stein}'s lemma,'' \emph{J. Am.
  Stat. Assoc}, vol.~87, no. 420, pp. 1025--1039, 1992.

\bibitem{LuL:18}
\BIBentryALTinterwordspacing
Y.~M. Lu and G.~Li, ``Phase transitions of spectral initialization for
  high-dimensional nonconvex estimation,'' \emph{Information and Inference},
  2018. [Online]. Available: \url{https://arxiv.org/abs/1702.06435}
\BIBentrySTDinterwordspacing

\bibitem{MondelliM:17}
M.~Mondelli and A.~Montanari, ``Fundamental limits of weak recovery with
  applications to phase retrieval,'' \emph{arXiv:1708.05932}, 2018.

\bibitem{BaiY:12}
Z.~D. Bai and J.~Yao, ``On sample eigenvalues in a generalized spiked
  population model,'' \emph{Journal of Multivariate Analysis}, vol. 106, no.
  167--177, 2012.

\bibitem{sampford1953some}
M.~R. Sampford, ``Some inequalities on mill's ratio and related functions,''
  \emph{The Annals of Mathematical Statistics}, vol.~24, no.~1, pp. 130--132,
  1953.

\end{thebibliography}

\end{document}